\documentclass[journal,twoside]{IEEEtran}
 \pdfoutput=1
\usepackage{amsfonts}
\usepackage{cite}
\usepackage{amsmath}
\usepackage{extarrows}
\usepackage{bm}
\usepackage{booktabs}
\usepackage{enumerate}

\ifCLASSINFOpdf
   \usepackage[pdftex]{graphicx}
   \graphicspath{{../pdf/}{../jpeg/}}
   \DeclareGraphicsExtensions{.pdf,.jpeg,.png}
\else
   \usepackage[dvips]{graphicx}
   \graphicspath{{../eps/}}
   \DeclareGraphicsExtensions{.eps}
\fi

\begin{document}
\title{OAM Mode-Group Generation Method: \\Partial Arc Transmitting Scheme}

\author{Xiaowen~Xiong,~\IEEEmembership{Student Member,~IEEE,}
        Shilie~Zheng,~\IEEEmembership{Member,~IEEE,}
        Zelin~Zhu,
        Yuqi~Chen,~\IEEEmembership{Student Member,~IEEE,}
        Zhixia~Wang,
        Xianbin~Yu,~\IEEEmembership{Senior Member,~IEEE,} 
        \\Xiaofeng Jin
        and~Xianmin~Zhang,~\IEEEmembership{Member,~IEEE}
\thanks{This work was supported by the National Natural Science Foundation of China under Grant 61571391.}
\thanks{The authors are with the College of Information Science and Electronic Engineering, Zhejiang University, Hangzhou 310027, China (e-mail: zhengsl@zju.edu.cn).}
}


\maketitle

\begin{abstract}
A partial slotted curved waveguide leaky-wave antenna which can generate orbital angular momentum (OAM) mode-groups (MG) with high equivalent OAM order $l_{e}=\pm 40$ at 60 GHz is proposed in this paper. The proposed antenna with partial slotting is designed according to the circular traveling-wave antenna which can generate single conventional OAM wave, so it can be regarded as partial arc transmitting (PAT) scheme compared with the full $2\pi$ aperture slotting of the circular traveling-wave antenna. The full-wave simulation results show that the generated OAM MGs present a high gain beam with a helical phase distribution. This method may lead to novel applications for next generation communication and radar system. 
\end{abstract}

\begin{IEEEkeywords}
Orbital angular momentum, mode-group, high equivalent OAM order, leaky-wave antenna, partial arc transmitting scheme.
\end{IEEEkeywords}
%
\IEEEpeerreviewmaketitle

\section{Introduction}
\IEEEPARstart{E}{xcept} for spin angular momentum (SAM) which represents the polarization feature of electromagnetic (EM) wave, orbital angular momentum (OAM) as another inherent characteristic of EM describes the spatial phase distribution and can be considered as a new degree of freedom to be utilized\cite{allen1992orbital}. OAM has recently derived many potential applications. In addition to multiplexing in communication systems\cite{sasaki2018experiment}, OAM has already been applied to nanotechnology\cite{drevinskas2015femtosecond}, quantum information technique\cite{chen2014quantum} and radar computational imaging\cite{chen2017single}, \emph{etc}.

Theoretically, OAM has an infinite number of spatial orthogonal modes, hence, same-frequency OAM multiplexing is regarded as a potential technique to increase the spectral efficiency. However, from the point of space division multiplexing, several researchers believe that OAM multiplexing is just a subset of multiple-in-multiple-out (MIMO) technique\cite{edfors2011orbital,zhao2015capacity}. In nature, it is a matter of signal processing from the physical layer\cite{zhang2016mode}. Besides, for practical application scenarios, especially in the RF domain, OAM wave presents a ``doughnut'' intensity shape because of its inherent divergence and central phase singularity. Furthermore, the ``doughnut'' will expand with the increasing propagation distance, which leads to the receiving problems at the receiving end for both the communication systems and the radar systems\cite{liu2019oam}. The partial arc receiving method is a great idea to solve this shortcoming, but it needs the specific receiving aperture according to several specific OAM modes\cite{zhang2016orbital}. Another apparent approach is to achieve the beam focusing with the help of lens  \cite{chen2015flat}, but it cannot change the ``doughnut'' form. Once beyond a limited distance, it's still difficult to receive with a whole angular aperture.    

\begin{figure}[t]
	\centering
	\includegraphics[width=3.0in]{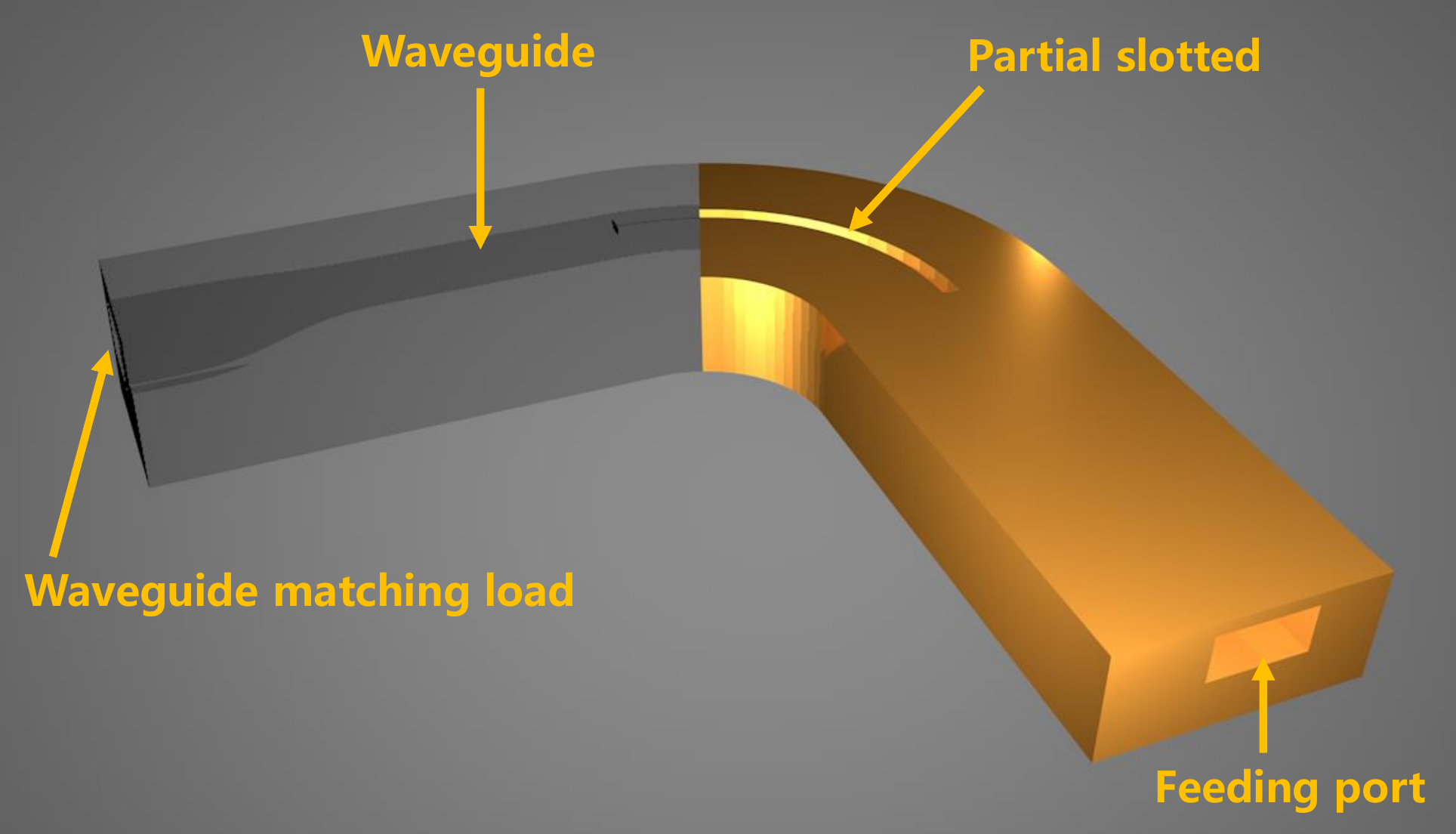}
	\caption{The diagrammatic drawing of partial slotted curved waveguide antenna.}
	\label{fig1}	
\end{figure} 
In order to reduce the impact of above-mentioned, it's highly necessary to realize a convergent beam which still retains a helical phase wavefront (a linear phase distribution along the azimuthal direction) locally. OAM mode-group (MG) has been proved that it's a field eigenmodes beamforming method to achieve radiation pattern diversity, it has notable characteristics including \textbf{\emph{directionality}}, \textbf{\emph{vorticity}} and \textbf{\emph{quasi-orthogonality}}\cite{ZhengRealization}. The prospects of OAM MGs in multiple-in-multiple-out (MIMO) system \cite{xiong2020performance}, spatial field digital modulation system\cite{chen2020ICCWS} and low interception communication system\cite{zhou2019low} have been explored. In \cite{xiong2020performance}, it's badly in need of a MG with higher equivalent OAM order $l_{e}$, where $l_{e}$ depicts the phase slope within the mainlobe, i.e. the \textbf{\emph{vorticity}} of a MG.  

The implemented method of OAM MGs in \cite{ZhengRealization} is to generate several OAM modes and then superpose them together into a MG, it's possible for low $l_{e}$. However, being aim at realizing a MG with higher $l_{e}$, it will be hard to generate several high order OAM modes, not to mention how to superpose them in practice. Azimuthal angle and OAM modes are connected with Fourier transformation, it has been proposed in optical domain that an angular restriction of light profile modifies the OAM spectrum, which means more OAM spectrum components are generated and these components can form a MG \cite{jack2008angular}. In RF domain, referring to the concept in \cite{jack2008angular}, researchers utilize the quasi-circular array\cite{fouda2018quasi} and density-weighted circular array\cite{liu2019oam} to realize one kind distinctive OAM wave which has a helical wavefront but a convergence energy distribution. However, the $l_{e}$ of generated beam in \cite{liu2019oam} and \cite{fouda2018quasi} are both low. Essentially, they can be regarded as an array amplitude-phase controlling method.

In this paper, OAM MGs with high equivalent OAM order $l_{e}=\pm40$ are generated by a partial slotted curved waveguide, which is a kind of leaky-wave antenna. The curved waveguide is designed as a part arc of the circular traveling-wave slot antenna which can generate single OAM mode\cite{zheng2015transmission}, so the proposed method is known as a partial arc transmitting (PAT) scheme. Different from the excitation way by a $90^{\circ}$ hybrid coupler in \cite{hui2015multiplexed}, this antenna is excitated from one port using coaxial adapter to waveguide, the other port is connected with a waveguide matching load, so a traveling-wave is formed in the curved waveguide, Fig. \ref{fig1} illustrates the sketch map of the proposed antenna. The antenna model is designed and simulated using full-wave analysis software CST Studio Suite 2018. This antenna can be applied to a wireless Line-of-sight MIMO link to achieve a longer effective communication distance\cite{xiong2020performance}.    

\section{ANTENNA GEOMETRY}
\begin{table}[b]
	\begin{center}
		\caption{Parameters for the antenna}
		\begin{tabular}{c|c}
			\toprule[2pt]
			\textbf{Symbols}&\textbf{Size (mm)}\\
			\hline
			$r_{i}$: the inner radius of curved waveguide&78.60\\ 
			$a_{w}$: the wide wall of curved waveguide&2.80\\ 
			$b_{w}$: the narrow wall of curved waveguide&0.68\\ 
			$a_{f}$: the wide wall of waveguide feeding port&WR-15\\ 
			$b_{f}$: the narrow wall of waveguide feeding port&WR-15\\ 
			$h_{s}$: the width of slot&0.70\\
			\bottomrule[2pt]
		\end{tabular}
		\label{tab1}
	\end{center}
\end{table}
In optical domain, a restriction of the angular range within an OAM light beam  profile will modify the OAM spectrum and generate OAM sidebands around the center OAM order, which can be explained using the ``angular diffraction theory''\cite{jack2008angular}. Due to the paraxial approximate property in optical domain, the OAM spectrum of the transmitted light through the angular mask presents a typical $sinc$ envelope, i.e. a symmetrical distribution. The transmitted light beam show a linear phase distribution in the mainlobe. 

Drawing on the idea in \cite{jack2008angular}, in RF domain, the partial slotted curved waveguide leaky-wave antenna is proposed, which is designed based on the circular traveling-wave antenna \cite{zheng2015transmission} and can be regarded as part of it. A circular traveling-wave antenna with its phase change along the circle of $2\pi l$ can generate the OAM wave with OAM order $l$. Such an antenna has a full $2\pi$ aperture slotting to achieve the leaky-wave radiation. The $2\pi/9$ partial slotting of the proposed antenna is a way to achieve the angular restriction within the RF OAM beam profile, which also can be considered as a PAT scheme in RF domain. The PAT scheme could also modify the OAM spectrum and make the energy of the center OAM mode transfer to the OAM sidebands. The quasi-symmetrical OAM components form an OAM MG which shows a high gain beam with a linear phase distribution in the mainlobe. 
\begin{figure}[t]
	\centering
	\includegraphics[width=3.6in]{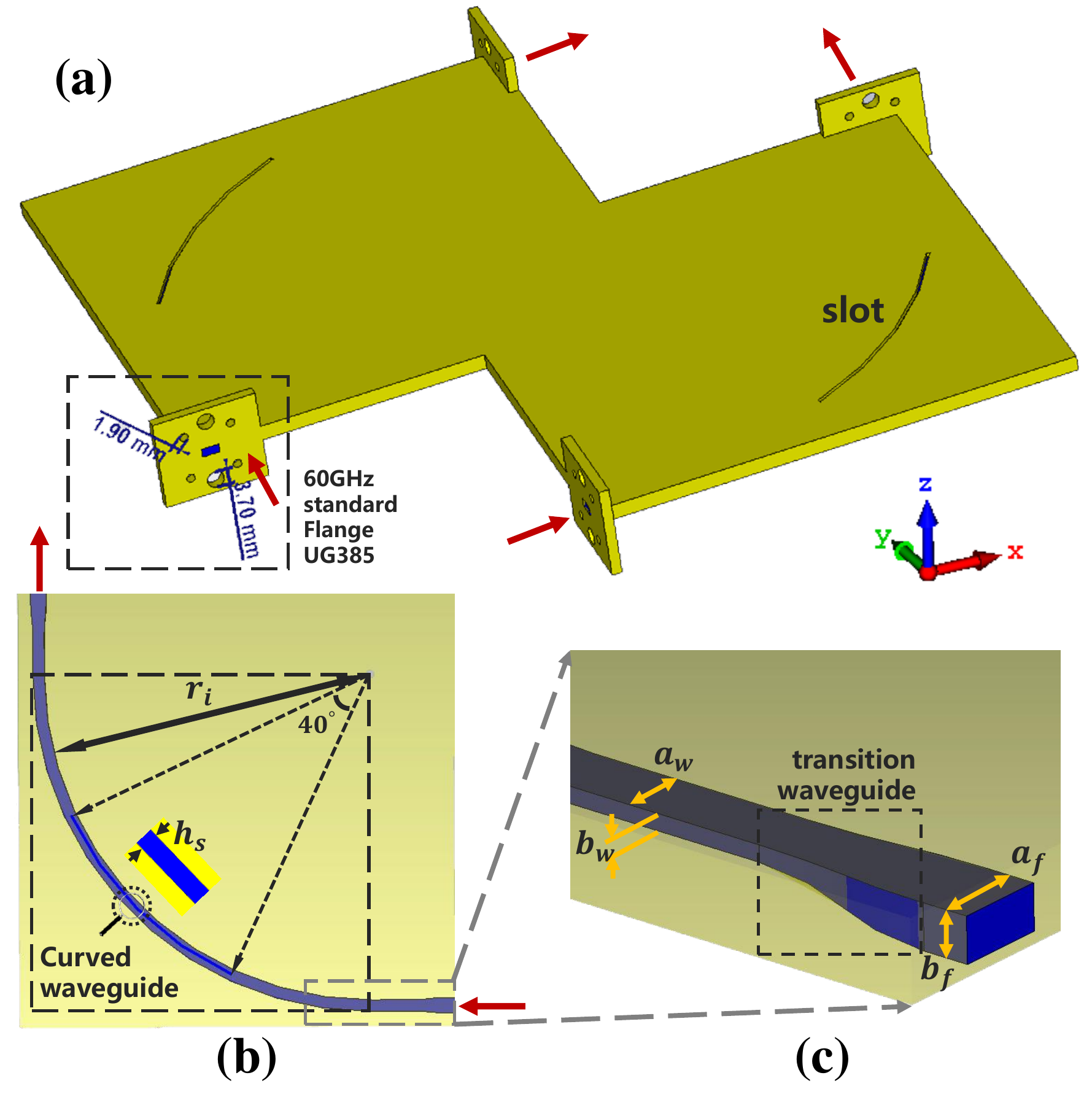}
	\caption{The antenna structure of the proposed PAT scheme. (a) Configuration of the partial slotted curved waveguide leaky-wave antenna. (b) Top views of the designed antenna, Table \ref{tab1} gives the significant design parameters. (c) Partial enlarged view of the ports, the transition waveguide is used for the matching between the waveguide feeding/terminal ports and the curved waveguide.}
	\label{fig2}	
\end{figure}
\begin{figure}[b]
	\centering
	\includegraphics[width=3.0in]{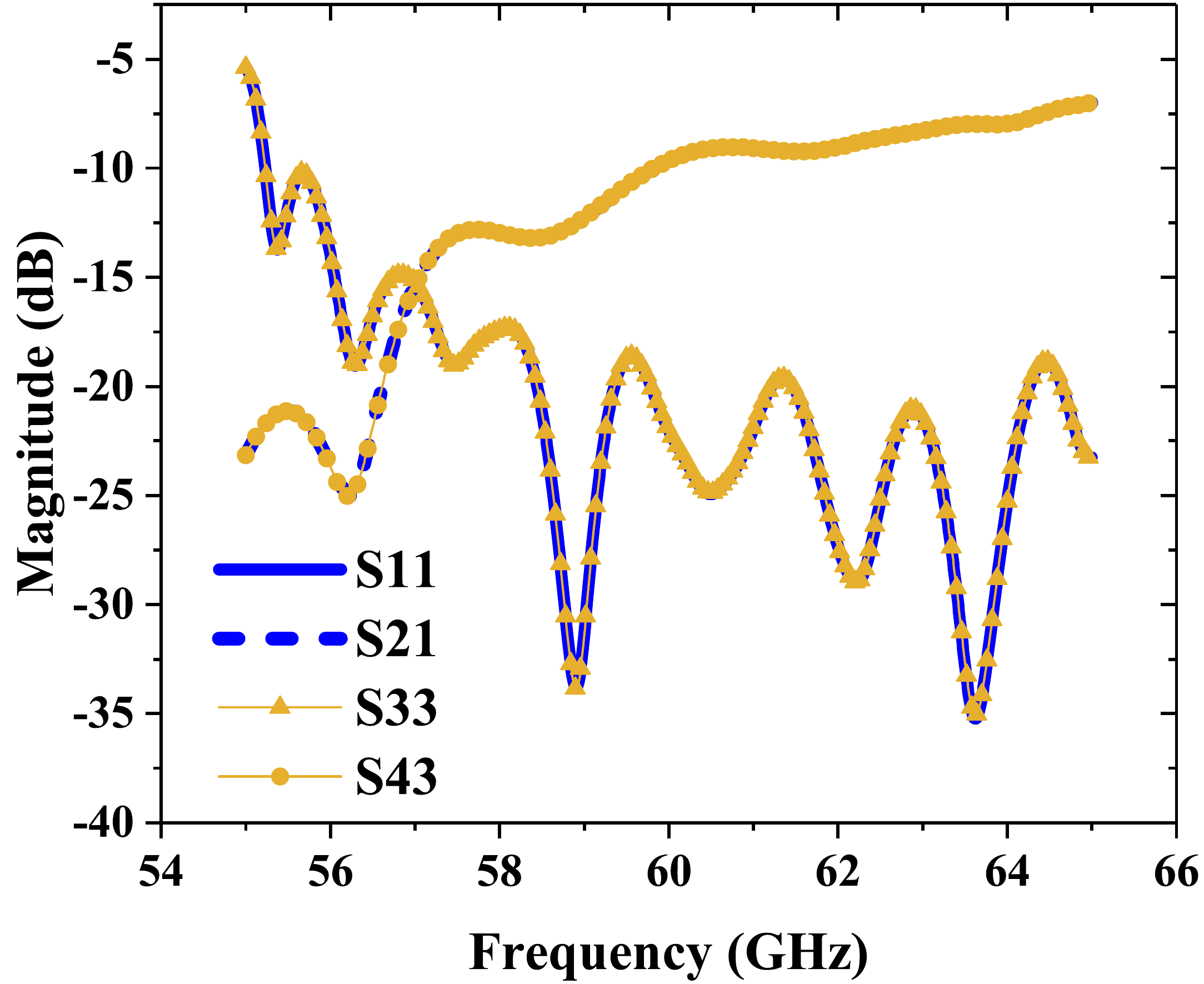}
	\caption{The simulated S-parameters of the proposed antenna. $S_{11}$ \& $S_{33}$: the return loss; $S_{21}$ \& $S_{43}$: the transmission loss.}
	\label{fig3}	
\end{figure}

A partial slotted curved waveguide leaky-wave antenna consists of the waveguide feeding/terminal matching port, the transition waveguide and the curved waveguide, Fig. \ref{fig2} (a) show the integral structure. The curved waveguide is fabricated by twisting a rectangular waveguide with a wide wall $a_{w}$ and a narrow wall $b_{w}$ into a $90^{\circ}$ arc as shown in Fig. \ref{fig2} (b). The energy is fed from one port using coaxial adapter to waveguide WR-15 (standard UG385 Flange). The other port is terminated with a waveguide matching load WR-15 (standard UG385 Flange), a traveling-wave is formed in the curved waveguide. A $40^{\circ}$ slot is cut on the wide wall, the OAM MG with high equivalent OAM order will radiate from it. The antenna is operated in the fundamental mode $\rm{TE}_{10}$, so $a_{w}$ should be less than $\lambda_{0}$ and greater than $\lambda_{0}/2$, where $\lambda_{0}$ is the wavelength in vacuum. Besides, the curved waveguide is designed according to the circular traveling-wave antenna, the circumference of circular traveling-wave antenna should be $l \lambda_{g}$, the waveguide wavelength $\lambda_{g}$ of $\rm{TE}_{10}$ can be calculated by
\begin{equation}
	\label{formula1}
	\lambda_{g\rm{TE}_{10}}=\frac{\lambda_{0}}{\sqrt{1-(\lambda_{0}/2a_{w})^2}}
\end{equation}      
the inner radius $r_{i}$ is equal to the inner radius of circular traveling-wave slot antenna, which is given by\cite{hui2015multiplexed}
\begin{equation}
	\label{formula2}
	r_{i}=\frac{|l|}{\pi\sqrt{(2/\lambda_{0})^2-(1/a_{w})^2}}-\frac{a_{w}}{2}
\end{equation}

In order to analyze the performance of the proposed scheme, an antenna model which can generate OAM MGs with high equivalent OAM order $l_{e}=\pm40$ at 60 GHz is simulated in full-wave analysis software. In this case, the significant design parameters have been given in Table \ref{tab1}. In fact, conditions permitting, more curved waveguides can be integrated on the same metal sheet to achieve more OAM MGs generating and multiplexing.   

\section{simulation results analysis}
\subsection{Antenna performance }
The S-parameters of the partial slotted curved waveguide leaky-wave antenna are shown in Fig. \ref{fig3}. It can be seen that it has a low reflection coefficient around 60 GHz owing to its traveling-wave structure, the waves above the cut-off frequency can be propagated in the curved waveguide. The difference of OAM equivalent order between OAM MGs ($l_{e}=\pm 40$) only depends on the direction of traveling-waves, as indicated in Fig. \ref{fig2} by the red arrows. The two curved waveguides are with the same size, besides, the width and the position of slots are also the same, so the S-parameters performance of both are almost uniform. Fig. \ref{fig3} shows the results of the OAM MG with $l_{e}=40$ (blue) and the OAM MG with $l_{e}=-40$ (yellow). 
\begin{figure}[t]
	\centering
	\includegraphics[width=3.4in]{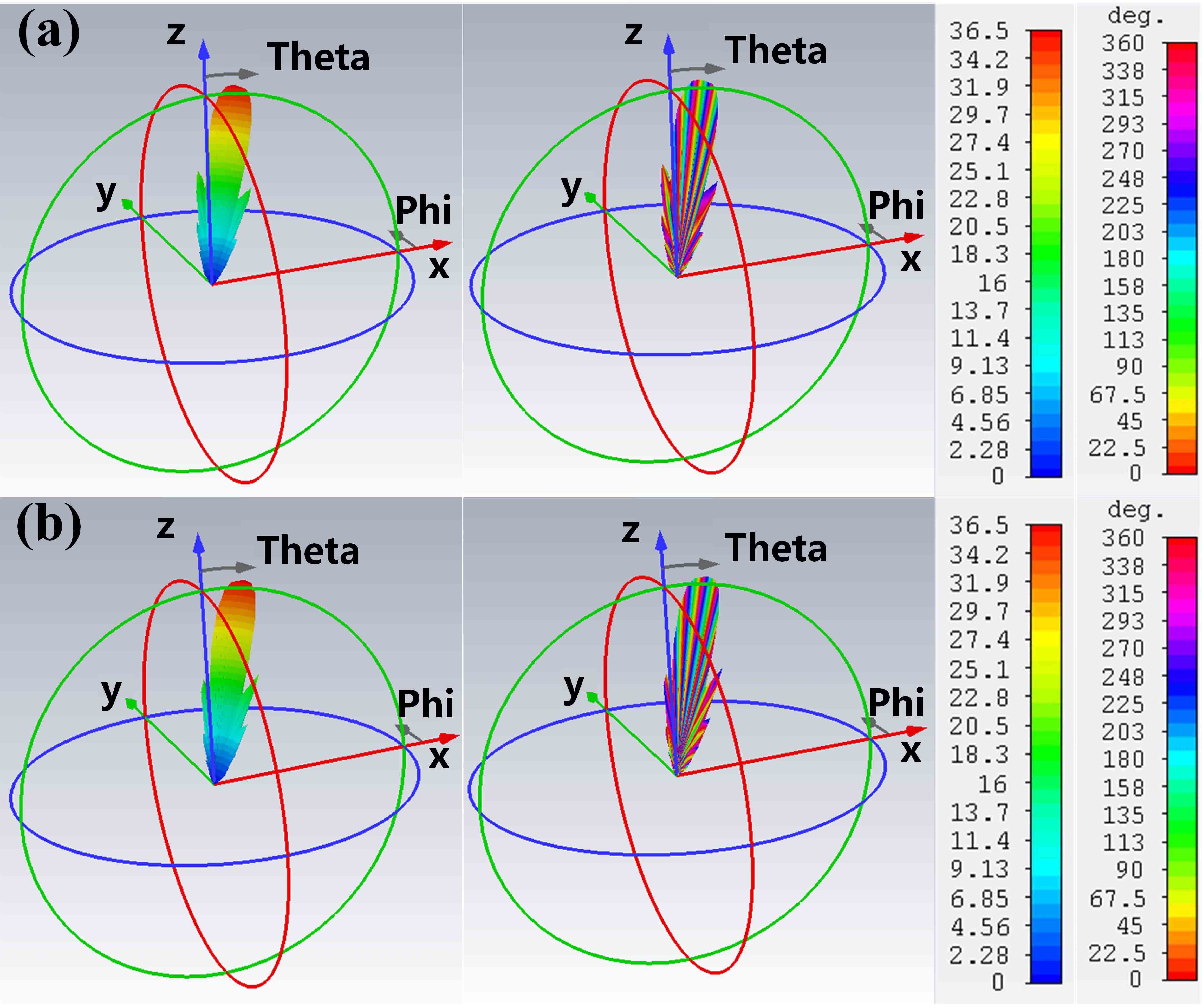}
	\caption{The simulated far-field 3D radiation pattern for the OAM MGs. (a) $l_{e}=40$; (b) $l_{e}=-40$.}
	\label{fig4}	
\end{figure}

The simulated far-field 3D radiation patterns of the proposed antenna is illustrated in Fig. \ref{fig4}. The divergence angle, 3 dB beam width of theta angle, directivity and radiant efficiency for $l_{e}=40$ are nearly $24.5^{\circ}$, $5.6^{\circ}$, 36.93 (15.67 dBi) and $85.25\%$, respectively. As for $l_{e}=-40$, these parameters are $24.7^{\circ}$, $5.5^{\circ}$,  37.3 (15.7 dBi) and $85.26\%$, respectively. The intensity pattern of two OAM MGs are almost the same, however, they show distinct-different phase distribution within the mainlobe.

\begin{figure}[t]
	\centering
	\includegraphics[width=3.5in]{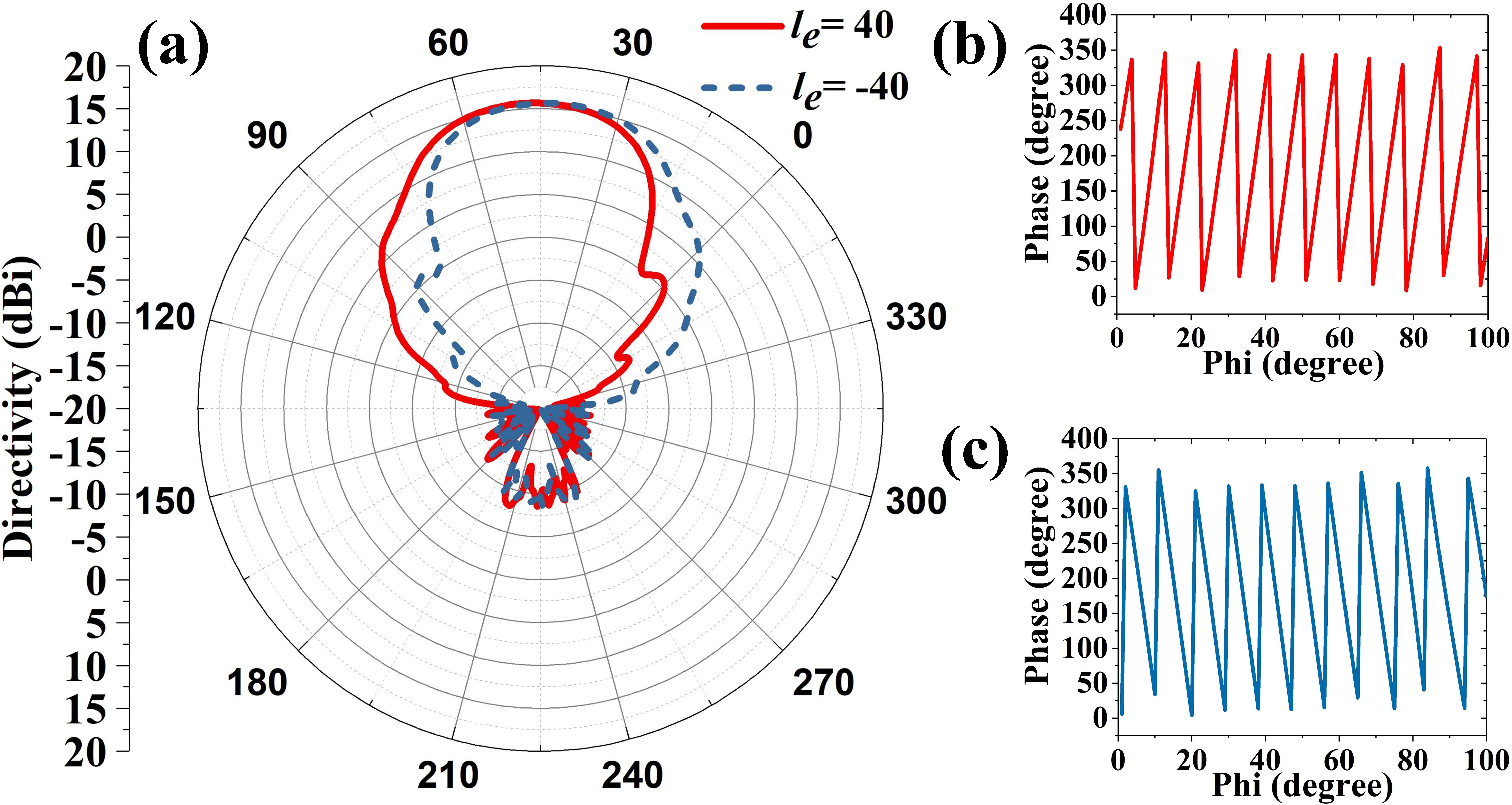}
	\caption{(a) The directivity for $l_{e}=\pm40$ at $\theta=24^{\circ}$ plane; (b) and (c) are the phase distribution within the mainlobe for $l_{e}=\pm40$ at $\theta=24^{\circ}$ plane.}
	\label{fig5}	
\end{figure}

In order to describe the azimuthal phase distribution more clearly, Fig. \ref{fig5} shows the directivity and phase distribution around azimuthal angle $\varphi$ within the OAM MG's mainlobe at $\theta=24^{\circ}$ plane. MGs inherit the helical phase distribution of conventional OAM-carrying waves, i.e. MGs still have the feature of \textbf{\emph{vorticity}}. The phase changes one period every $9^{\circ}$ azimuthal angle, which means the estimation of equivalent OAM order is $l_{e}=40$ or $l_{e}=-40$. The phase linearity of both cases are about good. To our knowledge, the data streams carried by the multiplexing OAM MGs can be demodulated by partial arc receiving method\cite{zhang2016orbital} utilizing the linear phase distribution. The mainlobe magnitude, mainlobe direction and 3 dB beam width of phi angle for $l_{e}=40$ are 15.7 dBi, $47^{\circ}$ and $44.6^{\circ}$, respectively. For $l_{e}=-40$, they are  15.6 dBi, $43^{\circ}$ and $47.5^{\circ}$, respectively.

\subsection{OAM spectrum analysis}
OAM modes among a MG and the azimuthal angle are connected by Fourier transformation. In optical domain, for the case of a restriction mask whose angle is $\gamma$ within the $l_{center}$th OAM light beam profile, the OAM spectrum of the transmitted light could be expressed as\cite{jack2008angular}
\begin{equation}
	\label{formula3}
	W(l)=sinc\left[\frac{\gamma(l-l_{center})}{2}\right]
\end{equation}   
where $W(l)$ refers to the weight of the $l$th OAM mode. Its OAM spectrum has a symmetrical weight distribution. Hence, the transmitted light demonstrates a helical phase distribution within its mainlobe, the phase slope is equal to $l_{center}$.
 
Without the paraxial approximate property, in RF domain, different OAM waves have inconsistent divergence angles. High order OAM wave has stronger divergence. The OAM spectrum of the MGs generated by the PAT scheme can be given by ``unpublished''\cite{ZhuICC}
\begin{equation}
	\label{formula4}
	W(l)=sinc\left[\frac{\varphi_s(l-l_{center})}{2}\right]J_{l}\left[ k r_s sin(\theta_0)\right] 
\end{equation} 
where $\varphi_s$ represents the angle corresponding to the arc length of slot, its value is $40^\circ$. $J_{l}(*)$ refers to the $l$th Bessel function of the first kind. $k$ is the wave number in free space. $r_s$ refers to the radius of the slot. $\theta_0$ is the divergence angle of the MGs.

\begin{figure}[t]
	\centering
	\includegraphics[width=3.5in]{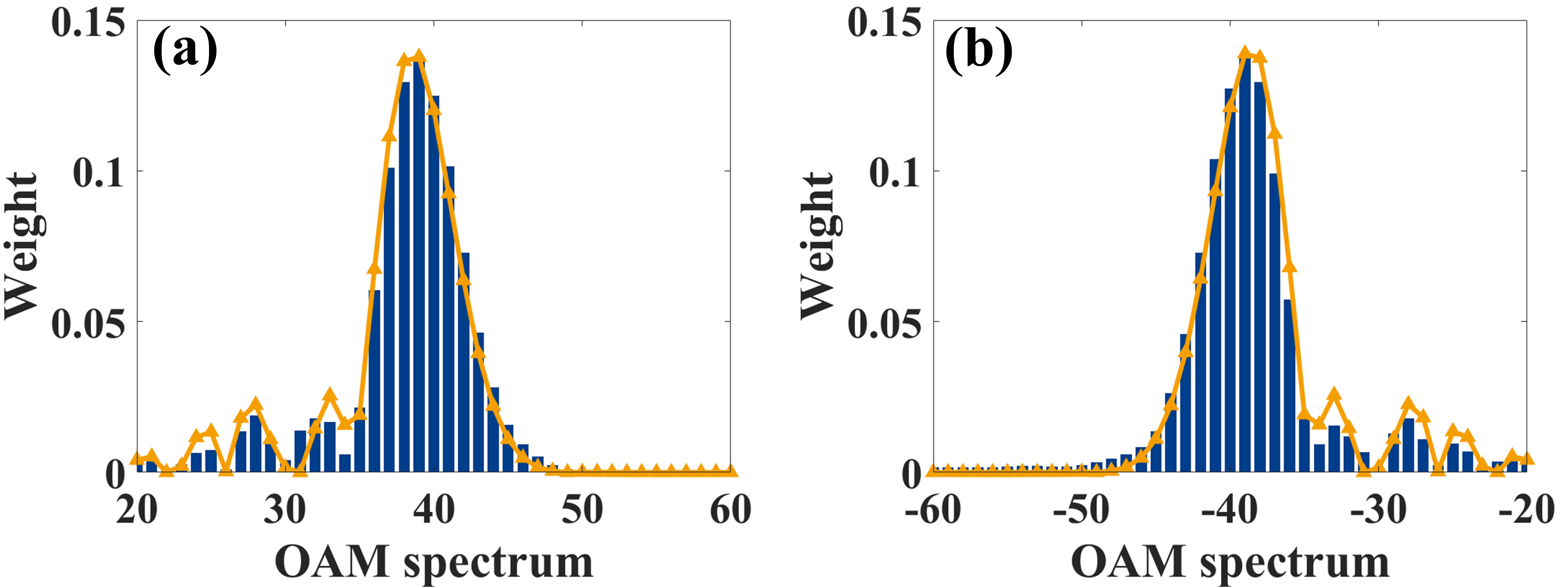}
	\caption{The OAM spectrum analysis of the simulated results (blue bar graph) and the theoretical results (yellow line), $\theta_0=24^\circ$. (a) $l_{e}=40$; (b) $l_{e}=-40$.}
	\label{fig6}	
\end{figure}

The OAM spectrum analysis of the simulated results and theoretical results are illustrated in Fig. \ref{fig6}. There is a good agreement between simulation and theory relatively. The symmetry of weight distribution is broken to some extent. High order OAM components are suppressed because of its severe divergence. The OAM spectrum components of MG $l_{e}=40$ are concentrated between $l=35$ and $l=45$, the weights nearly present a quasi-symmetrical distribution around the central mode $l_{center}=40$, which results in the good phase linearity of the MG within its mainlobe. As for MG $l_{e}=-40$, it shows a similar spectrum distribution. 

\section{Conclusion}
In this paper, we proposed a partial slotted curved waveguide leaky-wave antenna to directly generate OAM MGs with high equivalent OAM order $l_{e}=\pm40$ for 60 GHz operation. Such a partial arc transmitting (PAT) scheme can solve the problem that it's very hard to generate several high order OAM modes independently and superpose them into an OAM MG. The proposed method is much simpler for a practical system implementation. The simulated results show that the generated OAM MGs present high beam gain and helical phase distribution within the mainlobe. Our future work is going to focus on fabricating and measuring the antenna, then explore its potential application in the remote communication system \cite{XiongICC} and radar detection and positioning system.

\ifCLASSOPTIONcaptionsoff
  \newpage
\fi

\bibliographystyle{IEEEtran}
\bibliography{ref}

\begin{thebibliography}{10}
\providecommand{\url}[1]{#1}
\csname url@samestyle\endcsname
\providecommand{\newblock}{\relax}
\providecommand{\bibinfo}[2]{#2}
\providecommand{\BIBentrySTDinterwordspacing}{\spaceskip=0pt\relax}
\providecommand{\BIBentryALTinterwordstretchfactor}{4}
\providecommand{\BIBentryALTinterwordspacing}{\spaceskip=\fontdimen2\font plus
\BIBentryALTinterwordstretchfactor\fontdimen3\font minus
  \fontdimen4\font\relax}
\providecommand{\BIBforeignlanguage}[2]{{%
\expandafter\ifx\csname l@#1\endcsname\relax
\typeout{** WARNING: IEEEtran.bst: No hyphenation pattern has been}%
\typeout{** loaded for the language `#1'. Using the pattern for}%
\typeout{** the default language instead.}%
\else
\language=\csname l@#1\endcsname
\fi
#2}}
\providecommand{\BIBdecl}{\relax}
\BIBdecl

\bibitem{allen1992orbital}
L.~Allen, M.~W. Beijersbergen, R.~Spreeuw, and J.~Woerdman, ``Orbital angular
  momentum of light and the transformation of laguerre-gaussian laser modes,''
  \emph{Physical Review A}, vol.~45, no.~11, p. 8185, 1992.

\bibitem{sasaki2018experiment}
H.~Sasaki, D.~Lee, H.~Fukumoto, Y.~Yagi, T.~Kaho, H.~Shiba, and T.~Shimizu,
  ``Experiment on over-100-gbps wireless transmission with oam-mimo
  multiplexing system in 28-ghz band,'' in \emph{2018 IEEE Global
  Communications Conference (GLOBECOM)}.\hskip 1em plus 0.5em minus 0.4em\relax
  IEEE, 2018, pp. 1--6.

\bibitem{drevinskas2015femtosecond}
R.~Drevinskas, M.~Gecevi{\v{c}}ius, M.~Beresna, and P.~G. Kazansky,
  ``Femtosecond laser nanostructuring for high-topological charge vortex
  tweezers with continuously tunable orbital angular momentum,'' in \emph{The
  European Conference on Lasers and Electro-Optics}.\hskip 1em plus 0.5em minus
  0.4em\relax Optical Society of America, 2015, p. CLEO\_ECBO\_2\_3.

\bibitem{chen2014quantum}
L.~Chen, J.~Lei, and J.~Romero, ``Quantum digital spiral imaging,''
  \emph{Light: Science \& Applications}, vol.~3, no.~3, p. e153, 2014.

\bibitem{chen2017single}
Y.~Chen, S.~Zheng, X.~Jin, H.~Chi, and X.~Zhang, ``Single-frequency
  computational imaging using oam-carrying electromagnetic wave,''
  \emph{Journal of Applied Physics}, vol. 121, no.~18, p. 184506, 2017.

\bibitem{edfors2011orbital}
O.~Edfors and A.~J. Johansson, ``Is orbital angular momentum (oam) based radio
  communication an unexploited area?'' \emph{IEEE Transactions on Antennas and
  Propagation}, vol.~60, no.~2, pp. 1126--1131, 2011.

\bibitem{zhao2015capacity}
N.~Zhao, X.~Li, G.~Li, and J.~M. Kahn, ``Capacity limits of spatially
  multiplexed free-space communication,'' \emph{Nature photonics}, vol.~9,
  no.~12, p. 822, 2015.

\bibitem{zhang2016mode}
W.~Zhang, S.~Zheng, X.~Hui, R.~Dong, X.~Jin, H.~Chi, and X.~Zhang, ``Mode
  division multiplexing communication using microwave orbital angular momentum:
  An experimental study,'' \emph{IEEE Transactions on Wireless Communications},
  vol.~16, no.~2, pp. 1308--1318, 2016.

\bibitem{liu2019oam}
K.~Liu, Y.~Cheng, H.~Wang, and Q.~Yang, ``An oam-generating method using
  density-weighted circular array,'' in \emph{2019 20th International Radar
  Symposium (IRS)}.\hskip 1em plus 0.5em minus 0.4em\relax IEEE, 2019, pp.
  1--6.

\bibitem{zhang2016orbital}
W.~Zhang, S.~Zheng, Y.~Chen, X.~Jin, H.~Chi, and X.~Zhang, ``Orbital angular
  momentum-based communications with partial arc sampling receiving,''
  \emph{IEEE Communications Letters}, vol.~20, no.~7, pp. 1381--1384, 2016.

\bibitem{chen2015flat}
Y.~Chen, S.~Zheng, Y.~Li, X.~Hui, X.~Jin, H.~Chi, and X.~Zhang, ``A flat-lensed
  spiral phase plate based on phase-shifting surface for generation of
  millimeter-wave oam beam,'' \emph{IEEE Antennas and Wireless Propagation
  Letters}, vol.~15, pp. 1156--1158, 2015.

\bibitem{ZhengRealization}
S.~Zheng, Y.~Chen, Z.~Zhang, X.~Jin, H.~Chi, X.~Zhang, and Z.~N. Chen,
  ``Realization of beam steering based on plane spiral orbital angular momentum
  wave,'' \emph{IEEE Transactions on Antennas and Propagation}.

\bibitem{xiong2020performance}
X.~Xiong, S.~Zheng, Z.~Zhu, X.~Yu, X.~Jin, and X.~Zhang, ``Performance analysis
  of plane spiral oam mode-group based mimo system,'' \emph{IEEE Communications
  Letters}, 2020.

\bibitem{chen2020ICCWS}
Y.~Chen, X.~Xiong, Z.~Zhu, S.~Zheng, and X.~Zhang, ``Orbital angular momentum
  mode-group based spatial field digital modulation: Coding scheme and
  performance analysis,'' in \emph{2020 IEEE International Conference on
  Communications Workshops (ICC Workshops)}.\hskip 1em plus 0.5em minus
  0.4em\relax IEEE, 2020, accepted.

\bibitem{zhou2019low}
J.~Zhou, S.~Zheng, X.~Yu, X.~Jin, and X.~Zhang, ``Low probability of intercept
  communication based on structured radio beams using machine learning,''
  \emph{IEEE Access}, 2019.

\bibitem{jack2008angular}
B.~Jack, M.~Padgett, and S.~Franke-Arnold, ``Angular diffraction,'' \emph{New
  Journal of Physics}, vol.~10, no.~10, p. 103013, 2008.

\bibitem{fouda2018quasi}
R.~M. Fouda, T.~C. Baum, and K.~Ghorbani, ``Quasi-orbital angular momentum
  (q-oam) generated by quasi-circular array antenna (qca),'' \emph{Scientific
  reports}, vol.~8, no.~1, p. 8363, 2018.

\bibitem{zheng2015transmission}
S.~Zheng, X.~Hui, X.~Jin, H.~Chi, and X.~Zhang, ``Transmission characteristics
  of a twisted radio wave based on circular traveling-wave antenna,''
  \emph{IEEE Transactions on Antennas and Propagation}, vol.~63, no.~4, pp.
  1530--1536, 2015.

\bibitem{hui2015multiplexed}
X.~Hui, S.~Zheng, Y.~Chen, Y.~Hu, X.~Jin, H.~Chi, and X.~Zhang, ``Multiplexed
  millimeter wave communication with dual orbital angular momentum (oam) mode
  antennas,'' \emph{Scientific reports}, vol.~5, no.~1, pp. 1--9, 2015.

\bibitem{ZhuICC}
Z.~Zhu, X.~Xiong, Y.~Chen, S.~Zheng, and X.~Zhang, ``Orbital angular momentum
  {Mode-Group} in radar system: Rotational speed measurement scheme,'' in
  \emph{IEEE ICC 2020 Workshop on Orbital Angular Momentum Transmission (IEEE
  ICC'20 Workshop - OAMT)}.

\end{thebibliography}

\end{document}